\documentclass[mnsc,nonblindrev]{informs4} 

\OneAndAHalfSpacedXI

\definecolor{darkgreen}{rgb}{0,0.35,0}

\usepackage{subcaption}
\usepackage{mathrsfs}
\usepackage{amsmath,bm}
\usepackage{graphicx}
\usepackage{diagbox}
\usepackage{booktabs}
\usepackage{url}
\definecolor{DarkBlue}{rgb}{0,0.08,0.45}
\usepackage[backref = false, bookmarks, colorlinks = true, plainpages = false, citecolor = DarkBlue , urlcolor = DarkBlue, filecolor = DarkBlue, linkcolor = DarkBlue]{hyperref}
\usepackage{natbib}
\usepackage{multibib}
\newcites{EC}{References} 
\usepackage{multirow}

\usepackage{algorithm}
\usepackage{algpseudocode}

\usepackage{newunicodechar}
\usepackage[utf8]{inputenc}

\usepackage{natbib}
 \bibpunct[, ]{(}{)}{,}{a}{}{,}%
 \def\bibfont{\small}%

 \usepackage{pgfplots}
\usepgfplotslibrary{fillbetween}
\pgfplotsset{compat=1.13}

\TheoremsNumberedThrough     
\ECRepeatTheorems

\EquationsNumberedThrough    

\begin{document}
\newcommand{\abs}[1]{\left|  #1 \right| }
\newcommand{\brak}[1]{\left(#1\right)}    
\newcommand{\crl}[1]{\left\{#1\right\}}   
\newcommand{\edg}[1]{\left[#1\right]}     
\newcommand{\norm}[1]{\|#1\|}
\newcommand{\floor}[1]{\lfloor #1 \rfloor}

\newcommand{\cA}{{\mathcal A}}
\newcommand{\cB}{{\mathcal B}}
\newcommand{\cD}{{\mathcal D}}
\newcommand{\cF}{{\mathcal F}}
\newcommand{\cG}{{\mathcal G}}
\newcommand{\cH}{{\mathcal H}}
\newcommand{\cK}{{\mathcal K}}
\newcommand{\cL}{{\mathcal L}}
\newcommand{\cM}{{\mathcal M}}
\newcommand{\cR}{{\mathcal R}}
\newcommand{\cS}{{\mathcal S}}
\newcommand{\cT}{{\mathcal T}}
\newcommand{\cX}{{\mathcal X}}
\newcommand{\cP}{{\mathcal P}}
\newcommand{\cV}{{\mathcal V}}

\newcommand{\mA}{{\mathbb A}}
\newcommand{\mV}{{\mathbb V}}
\newcommand{\mC}{{\mathbb C}}
\newcommand{\mR}{{\mathbb R}}
\newcommand{\mE}{{\mathbb E}}
\newcommand{\mw}{{\mathbb w}}
\newcommand{\mT}{{\mathbb T}}

\newcommand{\bb}{{\mathbf b}}
\newcommand{\bd}{{\boldsymbol d}}
\newcommand{\by}{{\mathbf y}}
\newcommand{\bI}{{\mathbf I}}
\newcommand{\bp}{{\mathbf p}}
\newcommand{\bc}{{\mathbf c}}
\newcommand{\bg}{{\mathbf g}}
\newcommand{\bl}{{\mathbf l}}
\newcommand{\bbf}{{\mathbf f}}
\newcommand{\bq}{{\mathbf q}}

\newcommand{\bcD}{{\boldsymbol{\mathcal{D}}}}
\newcommand{\bbD}{{\boldsymbol{\mathscr{D}}}}

\newcommand{\bx}{{\boldsymbol x}}
\newcommand{\bA}{{\mathbf A}}
\newcommand{\bB}{{\mathbf B}}
\newcommand{\bC}{{\mathbf C}}
\newcommand{\bD}{{\mathbf D}}
\newcommand{\bG}{{\mathbf G}}
\newcommand{\bL}{{\mathbf L}}
\newcommand{\bS}{{\mathbf S}}
\newcommand{\bQ}{{\boldsymbol Q}}
\newcommand{\bU}{{\mathbf U}}
\newcommand{\bV}{{\boldsymbol V}}
\newcommand{\bK}{{\boldsymbol K}}
\newcommand{\bX}{{\mathbf X}}
\newcommand{\bZ}{{\mathbf Z}}
\newcommand{\bF}{{\mathbf F}}

\newcommand{\bmu}{{\boldsymbol \mu}}
\newcommand{\bomega}{{\boldsymbol \omega}}
\newcommand{\bw}{{\boldsymbol w}}
\newcommand{\bW}{{\boldsymbol W}}
\newcommand{\balpha}{{\boldsymbol \alpha}}
\newcommand{\blambda}{{\boldsymbol \lambda}}
\newcommand{\bxi}{{\boldsymbol \xi}}

\newcommand{\btheta}{\boldsymbol{\theta}}
\newcommand{\bsigma}{\boldsymbol{\sigma}}
\newcommand{\bnu}{\boldsymbol{\nu}}
\newcommand{\bSigma}{\boldsymbol{\Sigma}}
\newcommand{\bgamma}{\boldsymbol{\gamma}}
\newcommand{\bs}{\boldsymbol{s}}
\newcommand{\bz}{\boldsymbol{z}}

\newcommand{\C}{\mathbb{C}}

\newcommand{\D}{\mathbb{D}}
\newcommand{\F}{\mathbb{F}}
\newcommand{\p}{\mathbb{P}}
\newcommand{\Q}{\mathbb{Q}}
\newcommand{\W}{\mathbb{W}}
\newcommand{\R}{\mathbb{R}}
\newcommand{\q}{\mathbb{Q}}

\newcommand{\tr}{{\rm tr}}

\newcommand{\id}{{\mathbbm 1}}

\newcommand{\expect}{\mathbb{E}}



\RUNTITLE{
}

\TITLE{Sample-Efficient ``Clustering and Conquer" for Parallel Large-Scale Ranking and Selection}


\ARTICLEAUTHORS{%
\AUTHOR{Zishi Zhang}
\AFF{Guanghua School of Management, Peking University, Beijing 100871, China, \EMAIL{zishizhang@stu.pku.edu.cn}}
\AUTHOR{Yijie Peng}
\AFF{School of Management and Engineering, Nanjing University, Nanjing 210008, China, \EMAIL{pengyijie@nju.edu.cn}}
} 

\ABSTRACT{%
This work aims to improve the sample efficiency of parallel large-scale ranking and selection (R\&S) problems by leveraging correlation information.
We modify the commonly used ``divide and conquer" framework in parallel computing by adding a correlation-based clustering step, transforming it into ``clustering and conquer". {Theoretically, we develop a novel gradient-based analysis framework and show that this seemingly simple modification substantially improves the performance of large-scale R\&S procedures. Our approach enjoys two key advantages: (1) it does not require highly accurate correlation estimation or precise clustering, and (2) it can be seamlessly integrated with various existing fixed-precision and fixed-budget R\&S procedures while achieving optimal sample complexity. We also introduce a new parallel clustering algorithm tailored to large-scale settings. Finally, in large-scale AI applications such as neural architecture search, our methods demonstrate superior performance.}

}%



\maketitle

%


\section{Introduction}

Ranking and selection (R\&S) aims to identify the best design from a finite set of alternatives, through conducting simulation and learning about their performances \citep{bechhofer1954single}. It is typically framed into two main formulations: fixed-precision and fixed-budget. Fixed-precision R\&S procedures terminate the simulation once a pre-specified level of precision is achieved. Notable examples include the stage-wise algorithm of \cite{rinott1978two} and the KN family \citep{kim2001fully, hong2005tradeoff, jeff2006fully}, among others. In contrast, fixed-budget R\&S procedures stop the simulation once a predetermined total simulation budget is exhausted, with the goal of optimizing precision. {This category includes algorithms such as OCBA \citep{chen2000simulation, fu2007simulation}, large-deviation-based approaches \citep{glynn2004large},
and expected value of information algorithms \citep{chick2001new,frazier2008knowledge}.} For a more comprehensive review of the R\&S literature, please refer to \cite{chen2015ranking} and \cite{hong2021review}.

In recent years, R\&S algorithms have found significant applications in AI-related problems, such as reinforcement learning \citep{zhu2024uncertainty} and Monte Carlo tree search \citep{liu2024efficient}. This growing body of work has led to a shift in research focus toward large-scale R\&S problems, especially in parallel computing environments \citep{hunter2017parallel}. Here, ``large-scale'' refers to a large number of alternatives, denoted by $p$.  
The literature on large-scale R\&S can generally be divided into two branches. One branch focuses on addressing the challenges related to parallel computing implementation, such as information communication, synchronization, and workload balancing. Notable works in this branch include \citet{luo2015fully} and \cite{ni2017efficient}. The other branch focuses on improving sample efficiency. Recent efforts in this branch primarily aim to modify the inefficient all-pairwise comparison paradigm of classic fully-sequential R\&S algorithms such as the KN family. 
For example, \cite{zhong2022knockout} introduce a Knockout Tournament (KT) paradigm (and its fixed-budget version, FBKT \citep{hong2022solving}), which restricts comparisons to ``matches" involving only two alternatives at a time. The PASS paradigm of \cite{pei2022parallel} compares each alternative against a common standard to avoid exhaustive pairwise comparisons. 

However, existing R\&S algorithms typically assume independence across alternatives, thereby discarding valuable shared information \citep{eckman2022posterior}.
Leveraging such shared information offers a promising direction for improving sample efficiency, yet this remains largely unexplored in the large-scale R\&S literature. Existing efforts in utilizing shared information can be broadly categorized into two streams. The first involves using additional contextual or covariate information \citep{l2019gaussian,shen2021ranking,du2024contextual}. However, this approach relies on specific problem structures and is not applicable for general-purpose use. The second stream focuses on exploiting correlation or similarity information \citep{fu2007simulation,frazier2009knowledge,qu2015sequential,zhou2023sequential}. However, none of these algorithms is fully suitable for large-scale problems, as they typically require precise estimation of correlation or similarity parameters.

Before introducing our algorithm, we first provide some necessary background: most prominent parallel R\&S procedures adopt the \textit{divide and conquer} framework \citep{ni2017efficient,zhong2022knockout}, where alternatives are randomly distributed across processors, then the local best is selected from each processor, and the global best is selected from these local bests. 
In our work, we extend the traditional \textit{divide and conquer} framework by adding a correlation-based clustering step. This modified approach, termed the \textit{Parallel Correlation} {\textit{Clustering and Conquer}} (P3C) procedure, clusters alternatives based on their correlation and assigns alternatives from the same cluster to a single processor, rather than distributing them randomly. Both theoretical and empirical evidence show that this simple modification significantly improves sample efficiency. Moreover, P3C enjoys the advantage of not requiring precise estimation of the correlation parameter, as it is sufficient to merely identify which alternatives are highly correlated. Other notable alternative assignment strategies for \textit{divide and conquer} framework include the seeding approaches \citep{hong2022solving,li2024surprising} and the referencing approach \citep{zhong2025reference}.
 Moreover, we note that this work studies a general setting without distinguishing the specific sources of correlation.
{Correlation among alternatives may arise from common random numbers
(CRN) used in simulation experiments \citep{chen2012effects}, or from
shared data used to evaluate their performance. For example, in neural
architecture search, evaluating different architectures on the same
sampled training or validation data can induce correlated performance
estimates.}

{The intuition behind P3C is that, \textit{clustering strongly positively
correlated alternatives together can effectively cancel out stochastic
fluctuations in the same direction.} Throughout this paper, we focus on positive correlations between
alternatives; thus, ``highly correlated'' refers to a large positive
correlation coefficient.
To be more specific, the reason why R\&S requires a large number of samples is due to the random simulation outputs, which occasionally leads to undesirable situations where ``good" alternatives may unexpectedly perform worse than ``bad" ones. If strongly correlated alternatives are grouped together for comparison, their fluctuations tend to align in the same direction to some extent. When a ``good" alternative occasionally performs poorly, the ``bad" alternatives are likely to show similar declines; conversely, when a ``bad" alternative performs better than expected, the ``good" alternatives tend to exceed expectations as well. This way, the true ranking is preserved, enabling us to identify the true best alternative with fewer simulations.
This concept is similar to the CRN technique, which introduces positive correlation artificially to reduce variance and expedite pairwise comparisons.
However, it is important to note that our theoretical analysis is fundamentally different from CRN, as we focus on the global impact of correlation information: for instance, one interesting conclusion from our analysis is that increasing the correlation between a pair of alternatives can increase the probability of selecting other ``good" alternatives that are \textit{not} directly related to this pair.}

To formalize the intuition of P3C, we develop a novel gradient analysis framework. In the literature, the probability of correct selection (PCS) is commonly used as a measure of the precision of R\&S procedures. It is defined as the probability that the sample average of the ``true best" alternative is higher than that of the others. We generalize the classical PCS by replacing the ``true best" with any alternative \(\tau\), termed as the individual PCS(\(\tau\)), which serves as a probabilistic criterion for assessing the performance of alternative \(\tau\). We then analyze the derivative of individual PCS with respect to correlation information to explore a novel ``{mean-covariance}" interaction, as opposed to the widely discussed ``mean-variance” tradeoff. {The analysis shows that, increasing correlation between alternatives induces an interesting ``separation" effect, which probabilistically amplifies good alternatives while suppressing the bad ones. 
These theoretical insights explain how correlation-based clustering can improve R\&S performance. Under a stylized symmetric benchmark, we further quantify the resulting PCS improvement under a fixed sample allocation. More broadly, the gradient analysis developed in this paper provides a new perspective on how correlation affects selection probabilities in general R\&S problems.
Similarly, a related work by \cite{peng2017gradient} also explores gradient analysis in R\&S, but it investigates the impact of ``induced correlation", which is induced by variance under the independent assumption, rather than the actual correlation existing between alternatives. Additionally, other works exploring gradient analysis in R\&S include \cite{peng2015non} and \cite{zhang2023gradient}.}

In recent literature, analyzing the asymptotic behavior of the required total sample size as \( p \to \infty \) has become a critical approach for assessing the sample efficiency of large-scale R\&S procedures \citep{zhong2022knockout,hong2022solving,li2024surprising}. As \( p \to \infty \), the theoretical lowest growth rate of the required total sample size to achieve non-zero precision asymptotically is \( \mathcal{O}(p) \), and a R\&S procedure that achieves this \( \mathcal{O}(p) \) sample complexity is referred to as \textit{sample-optimal}. Known sample-optimal procedures include the median elimination (ME) procedure \citep{even2006action}, the KT procedure and its fixed-budget version FBKT, as well as the greedy procedure in \cite{li2024surprising}.
As $p$ grows large, the proposed P3C performs multiple rounds of ``clustering and conquer”, resembling the knockout tournament scheme, and can also achieve sample optimality when combined with classic fixed-budget and fixed-precision R\&S algorithms.

{
The main contributions of this paper are as follows. In Section~\ref{chap_3_P3C}, we propose the P3C procedure. In Section~\ref{chap4}, we establish the sample optimality of P3C under both the fixed-precision and fixed-budget formulations and develop a novel gradient-based framework for analyzing the effects of correlation on PCS. This framework reveals a key mean--correlation interaction and quantifies the PCS improvement achieved by correlation-based clustering under a stylized symmetric benchmark with a fixed sample allocation. In Section~\ref{chap6}, we introduce a parallelizable few-shot clustering algorithm.}

\section{Problem Formulation}
Let $\mathcal{P}=\{1,2,\ldots,p\}$ denote the index set for all $p$ alternatives. We adopt a frequentist framework, and the output of alternative $i\in\mathcal{P}$ is a random variable $X_i$. We assume that the population distribution of the random vector $(X_1,X_2,\ldots,X_p)$ is multivariate normal $N(\bm{\mu},\Sigma_{p\times p})$, where $\bm{\mu}=(\mu_1,\cdots,\mu_p)$ is the mean vector and $\Sigma_{p\times p}$ is the covariance matrix. 
Let $x_{ij}$ denote the $j$th simulation observation of alternative $i$. The observation vectors $(x_{1j},x_{2j},\cdots,x_{pj}) \sim N(\bm{\mu},\Sigma_{p\times p})$ are independently and identically distributed. We assume that the covariance between \( x_{im} \) and \( x_{jn} \) is given by \( \mathrm{cov}( x_{im}, x_{jn} ) = 0 \) if \( m \neq n \) and \( \mathrm{cov}( x_{im}, x_{jn} ) = \mathrm{cov}(X_i, X_j) \) if \( m = n \), where \( \mathrm{cov}(X, Y) \) denotes the covariance between the random variables \( X \) and \( Y \).

Let $N_i$ denote the total sample size allocated to alternative $i$ and ${\bar{x}}_i$ be the sample average. 
Then $$\mathrm{cov}\left({\bar{x}}_i,{\bar{x}}_j\right)=\frac{\mathrm{cov}\left(X_i,X_j\right)}{N_{i,j}},$$ where $N_{i,j}\triangleq\max (N_i,N_j)$.  The correlation among alternatives may arise from any form of correlated sampling. In practice, some alternatives are more correlated than others, leading to a natural clustering structure. 
{We assume that the $p$ alternatives come from $k$ non-overlapping clusters based on correlation: $\mathcal{G}_1, \ldots, \mathcal{G}_k$, with cardinality $|\mathcal{G}_j|$. $G$ is a mapping from $\mathcal{P}$ to $\{1,2,\ldots,k\}$, where $G\left(i\right)=j$ if alternative $i$ belongs to $\mathcal{G}_j$. 
$\Pi=(G\left(1\right),G\left(2\right),\ldots,G(p))\in\mathbb{R}^p$ represents the true cluster partition.  The independent case is also included by setting $p=k$.  Alternatives within the same cluster are more highly correlated than those across clusters.}
The correlation structure, as well as the underlying cluster partition, is unknown and needs to be estimated using observed data.

Let $[m]$ denote the index of the alternative with $m$-largest mean, i.e., $\mu_{[1]}>\mu_{[2]}>\cdots>\mu_{[p]}$.
The objective of R\&S is to identify the unique true best alternative $$\left[1\right]=\mathop{\arg\max}\limits_{i\in \mathcal{P} }\ \mu _i,$$ under either a fixed-budget constraint or a fixed-precision constraint. Since the selection policy is typically specified as selecting  $\arg\max_{i\in \mathcal{P}}{\bar{x}}_i$, the corresponding precision metric is the traditional PCS, defined as $$\text{PCS}_{\text{trad}}\triangleq P\left({\bar{x}}_{[1]}>{\bar{x}}_j,j\neq[1]\right).$$ In this work, to facilitate a more general theory, for any alternative $\tau\in\mathcal{P}$, we define the individual PCS of $\tau$ as $$\text{PCS}\left(\tau\right)\triangleq P\left({\bar{x}}_\tau>{\bar{x}}_j,j\neq\tau\right),$$ which naturally generalizes the traditional PCS as $\text{PCS}_{\text{trad}}=\text{PCS}\left([1]\right)$. According to \cite{hong2021review}, the statistical meaning of $\text{PCS}\left(\tau\right)$ is the probability of rejecting the null hypothesis that ``$\tau$ is \textit{not} the true best", making it a useful metric for evaluating the performance of $\tau$.

 Recently, analyzing the asymptotic behavior as $p\to\infty$ has emerged as a critical approach for assessing the theoretical performance of large-scale R\&S procedures \citep{zhong2022knockout,hong2022solving,li2024surprising}. {This line of research has been developed under both the fixed-precision and fixed-budget formulations. Under the fixed-precision formulation, a procedure is sample optimal if it guarantees a prescribed PCS with an expected total sample size of $O(p)$. Under the fixed-budget formulation, sample optimality requires the rate for maintaining correct selection to be $\Theta(p)$; that is, a budget growing linearly in $p$ is sufficient to keep the probability of correct selection bounded away from zero. }
  
\section{The Framework of P3C: From ``Divide and Conquer" to ``Clustering and Conquer"}\label{chap_3_P3C}


``Divide and conquer" has long served as the foundational framework for many mainstream parallel R\&S procedures. In this framework, alternatives are randomly distributed across different processors, where each processor selects its local best alternative, and the global best is then selected from these local bests. We extend this framework by introducing a ``clustering and conquer" strategy, formally termed \textit{Parallel Correlation Clustering and Conquer} (P3C). The key distinction in P3C is an additional step of correlation-based clustering, which groups highly correlated alternatives and assigns each cluster to a single processor, as opposed to random assignment. This simple modification leads to a significant improvement in sample efficiency.

\begin{figure}
{
\centering
\includegraphics[width=1\textwidth]{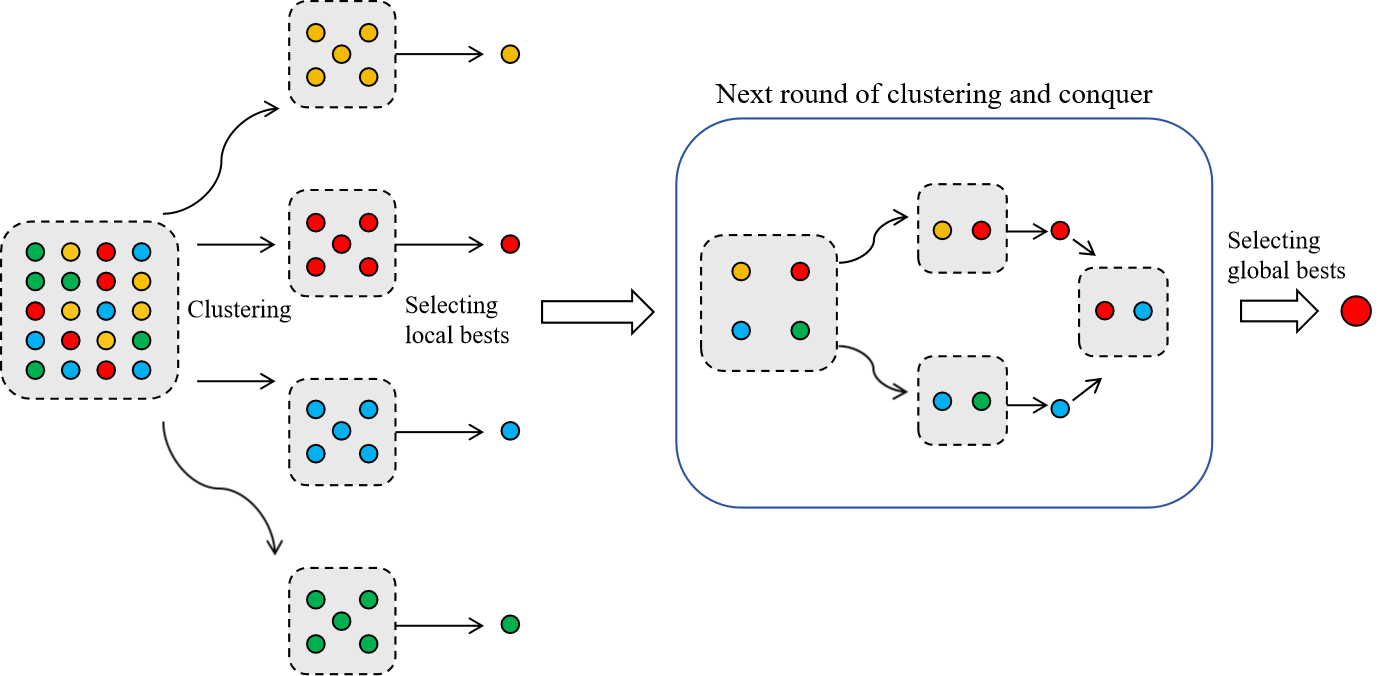}
\caption{Parallel correlation clustering and conquer.\label{fig:P3C}}
}
\end{figure}

Specifically, as outlined in Algorithm \ref{P3C} and Figure \ref{fig:P3C}, P3C starts with an initialization Stage 0. 
Then, in Stage 1, we continue sampling and then cluster the alternatives based on estimated correlations. The sample size can be predetermined based on the required clustering accuracy. Additionally, the novel few-shot clustering algorithm $\mathcal{A}\mathcal{C}^+$, detailed in Section \ref{chap6}, enables efficient parallelization of both Stage 0 and Stage 1 in large-scale problems. 
Upon completion of Stage 1, the alternatives of the same cluster are sent to a single processor. 

{
During Stage 2, we perform R\&S to select the \textit{local best} within each cluster. The choice of R\&S algorithm depends on the problem formulation. Under fixed precision, P3C-KN applies KN directly to the alternatives in each cluster, whereas P3C-KT runs a knockout tournament within each cluster, using KN to determine the winner of each pairwise match. Under fixed budget, P3C can use equal allocation (EA), OCBA, CBA, or other budget-allocation rules within each cluster.}

In Stage 3, we continue R\&S within these remaining \textit{local bests} to select the final winner.
However, when $p$ is large, the number of remaining \textit{local bests} after Stage 2 may still be substantial. In such cases, P3C employs repeated rounds of ``clustering and conquer" (i.e., Stages 1 and 2) until the number of remaining alternatives becomes manageable for a single processor. In practice, the maximum number of alternatives that a single processor can handle, denoted by \(p_m \geq 2\), is pre-estimated based on computational capacity.
\textcolor{blue}{
Clustering in different rounds are based on the initialization
samples collected before the selection procedures. These samples are used only
to estimate correlations and determine cluster assignments, and they are not
reused in subsequent R\&S procedures.
}

{In the fixed-precision R\&S, let the final desired precision be \(1 - \alpha\), where $\alpha$ is the error tolerance level. The error tolerance $\alpha_2^r$ for each processor in the \(r\)-th round of ``clustering and conquer", and the error tolerance $\alpha_3$ for Stage 3 in P3C must satisfy $\sum_r \alpha_2^r+\alpha_3=\alpha$.} We set $\alpha_2^r=\frac{\alpha-\alpha_3}{2^r}$.
This multi-round structure closely resembles the knockout tournament paradigm, and consequently, as shown in the following Section \ref{sec:fixed-precision-optimality}, P3C-KN and P3C-KT can also achieve sample optimality as \(p \to \infty\). 
{In the fixed-budget R\&S, P3C similarly divides the total budget $N$ across multiple rounds when $p$ is large.
Let $N^r$ denote the budget assigned to round $r$, with the budgets across all rounds summing to $N$.
In each round, $N^r$ is first distributed across the current clusters.
Within each cluster, its assigned budget can be allocated among the
alternatives using any fixed-budget R\&S procedure.}

Note that this paper focuses on analyzing the sample efficiency of the P3C, while many challenges in parallel implementation, such as master-worker task coordination, synchronization, and workload balancing, are beyond our scope. For comprehensive solutions to these issues, readers are encouraged to refer to the prior works of \cite{luo2015fully} and \cite{ni2017efficient}. It is also important to note that, in our setting, no additional structural assumptions are imposed, and highly correlated alternatives do not necessarily have similar means.
Therefore, grouping highly correlated alternatives does not exacerbate potential drawbacks associated with clustering mean-similar alternatives.
In addition, the clustering algorithm \(\mathcal{A}\mathcal{C}^+\) described in Section~\ref{chap6} incorporates specific steps to further prevent such cases.

\begin{algorithm}[htb]
\caption{Parallel Correlation Clustering and Conquer (P3C)}
\label{P3C}
\begin{algorithmic}[1] 
\State \textbf{Stage 0 (Initialization):} Simulate each alternative at least $N_0 \geq 3$ times and estimate the covariance matrix $\Sigma$.
\State \textbf{Stage 1 (Clustering):} Continue sampling and apply the correlation-based clustering algorithm $\mathcal{A}\mathcal{C}^+$ \textcolor{blue}{using the initialization samples collected in Stage 0} to partition all alternatives into several clusters. Assign all alternatives within the same cluster to a single processor.
\State \textbf{Stage 2 (Conquer):} Perform R\&S within each cluster to select the local best.
\State \textbf{Stage 3 (Final Comparison):} If the number of remaining alternatives in contention satisfies $\leq p_m$, perform R\&S to select the global optimal alternative. Otherwise, repeat Stage 1 and Stage 2 for the remaining alternatives.
\end{algorithmic}
\end{algorithm}

\section{Theoretical Analysis: Understanding Mean-Correlation Interactions}
\label{chap4}

{
This section studies the theoretical properties of P3C from two
perspectives. First, Section~\ref{sec:sample-optimality} establishes
the sample optimality of P3C in large-scale R\&S problems under both
the fixed-precision and fixed-budget formulations. This optimality
follows from P3C's multiround divide-and-conquer mechanism. Sample
optimality characterizes the asymptotic order, leaving room for
correlation-based clustering to improve PCS at finite sample sizes.
Second, Section~4.2 examines the additional benefit of
correlation-based clustering by focusing on the local selection
problem on one processor in one round. The analysis identifies a
``mean--correlation interaction'': stronger correlation tends to
increase the selection probabilities of high-mean alternatives while
decreasing those of low-mean alternatives. This interaction helps
explain how correlation-based clustering can improve PCS.
The two analyses consider distinct asymptotic regimes.
The sample-optimality analysis adopts the large-scale R\&S regime in
which $p\to\infty$.
The mean--correlation analysis instead holds
$p$ fixed and considers a prespecified deterministic sample
allocation. Corollary~\ref{co1} and the subsequent results examine
asymptotic orders as the deterministic total budget $N\to\infty$.
}

{\subsection{Sample Optimality of P3C}
\label{sec:sample-optimality}

This subsection establishes that P3C is sample optimal as the number $p$ of alternatives grows under both the fixed-precision and fixed-budget formulations. Following the large-scale asymptotic regime of \citet{li2024surprising}, we impose the following assumption.

\begin{assumption}
\label{assump:sample-optimality}
As $p\to\infty$, the identity of the unique best alternative $[1]$ remains unchanged, and there exist constants $\delta>0$, $\sigma_{\mathrm{upper}}^2<\infty$, and $p_m<\infty$, such that $\mu_{[1]}-\mu_{[2]}\geq\delta$, $\operatorname{Var}(X_i-X_j)\leq\sigma_{\mathrm{upper}}^2$ for all $i\neq j$, and the number of alternatives in each cluster is at most $p_m$.
\end{assumption}

\subsubsection{Fixed-precision formulation}
\label{sec:fixed-precision-optimality}
Fix a target precision $\alpha\in(0,1)$. According to
\citet{zhong2022knockout}, sample optimality in the fixed-precision formulation means
achieving $\mathrm{PCS}_{\mathrm{trad}}\geq1-\alpha$ with an expected total sample size
of $O(p)$ as $p\to\infty$. The following proposition establishes the sample optimality
of P3C-KN and P3C-KT. The proof is given in the appendix.

\begin{proposition}[Fixed-precision sample optimality]
\label{prop:fp-sample-optimality}
Suppose that Assumption~\ref{assump:sample-optimality} holds.  Let $N_{\mathrm{P3C-KN}}$ and $N_{\mathrm{P3C-KT}}$ denote the required total sample sizes to achieve $\mathrm{PCS_{trad}}\geq 1-\alpha$ using $\mathrm{P3C-KN}$ and $\mathrm{P3C-KT}$, respectively.  Then
\begin{equation*}
    \mathbb{E}\!\left[N_{\mathrm{P3C-KN}}\right]=O(p)
    \quad\text{and}\quad
    \mathbb{E}\!\left[N_{\mathrm{P3C-KT}}\right]=O(p)
    \label{eq:fp-linear-bound}
\end{equation*}
as $p\to\infty$. Consequently, both P3C-KN and P3C-KT are sample optimal under the
fixed-precision formulation.
\end{proposition}

\subsubsection{Fixed-budget formulation}
\label{sec:fixed-budget-optimality}
We next consider the fixed-budget formulation, in which the total simulation budget
$N$ is specified in advance. Following \citet{hong2022solving}, the \emph{rate for
maintaining correct selection} (RMCS) is the smallest order of $N$ required to keep the
probability of correct selection bounded away from zero as $p\to\infty$. A fixed-budget
procedure is sample optimal if its RMCS is $\Theta(p)$.

P3C can be combined with any budget-allocation rule within each cluster. We establish sample
optimality using equal allocation, which assigns the same number of observations
to every alternative and does not use information about their means, variances,
or correlations. EA therefore provides a basic and conservative benchmark for P3C.
More sophisticated allocation rules, such as OCBA and CBA, can also be used within the
P3C framework and may achieve substantially better finite-budget performance. More
generally, any within-cluster rule that retains the best alternative with conditional
probability no smaller than that under EA, given the same group budget and round
history, inherits the sample-optimality result established below.

Under EA, let $N^r$ denote the total budget allocated to round $r$. For a constant
$\phi\geq2$, we set
$N^r=\frac{r}{\phi(\phi-1)}\left(\frac{\phi-1}{\phi}\right)^rN$.
Because $\sum_{r=1}^{\infty}N^r=N$, the allocations over the finitely many rounds do
not exceed the total budget. Within each round, $N^r$ is allocated across clusters in
proportion to their sizes and then equally among the alternatives within each cluster. The proof is provided in the appendix.

\begin{proposition}[Fixed-budget sample optimality]
\label{prop:fb-sample-optimality}
Suppose that Assumption~\ref{assump:sample-optimality} holds and that EA is applied
within each cluster. For any constant $\eta_1\geq2\phi^2$, if $N>p\eta_1$, then
\begin{equation}
    \mathrm{PCS}_{\mathrm{trad}}
    \geq\exp\!\left(
        -\frac{p_m\pi^2\phi^2\sigma_{\mathrm{upper}}^2}
        {3\eta_1\delta^2}
    \right)>0,
    \label{eq:fb-pcs-lower-bound}
\end{equation}
where the lower bound is independent of $p$. Therefore, P3C-EA has RMCS $\Theta(p)$
and is sample optimal under the fixed-budget formulation.
\end{proposition}

Propositions~\ref{prop:fp-sample-optimality} and \ref{prop:fb-sample-optimality} establish that P3C is sample optimal under both large-scale fixed-precision and fixed-budget formulations. This sample optimality follows from the underlying multi-round divide-and-conquer mechanism and does not depend on the correlation-based clustering method or clustering quality. However, sample optimality characterizes only the asymptotic order, and procedures with the same order may still differ substantially in their leading constants and finite-sample performance. The following subsections show that, correlation-based clustering can further improve PCS by increasing the
correlations among alternatives assigned to the same processor.
}

\subsection{Interaction and Impact of Mean and Correlation on PCS}\label{sec4.1}

{
In this section, we develop a novel gradient-based analysis of how
correlation-based clustering affects R\&S performance in P3C.
Viewing
$\mathrm{PCS}(\tau;\bm\mu,\Sigma,\{N_i\}_{i\in\mathcal P})$
as a function of the mean vector $\bm\mu$, covariance matrix $\Sigma$,
and sample allocation $\{N_i\}_{i\in\mathcal P}$, we analyze its
derivatives to characterize how mean and correlation information
jointly affect selection probabilities.
The analysis reveals a ``separation'' effect: stronger correlation
tends to increase the selection probabilities of high-mean
alternatives while decreasing those of low-mean alternatives.
This effect explains why correlation-based clustering can improve PCS
in P3C.  We study this mechanism through the local selection problem
on one processor in one round. Because P3C performs such local
selections across processors and rounds, the resulting insights
extend naturally to the full procedure.

Our analysis based on the individual $\mathrm{PCS}(\tau)$ is general
and includes $\mathrm{PCS}_{\mathrm{trad}}$ as the special case $\tau=[1]$.
Its implications also extend beyond P3C and large-scale R\&S: the results
characterize how correlation, including that induced by CRN, affects selection accuracy in general R\&S problems.
To isolate the interaction between means and correlations, we consider
a fixed-budget R\&S setting with $p$ fixed and a deterministic sample
allocation $\{N_i\}_{i\in\mathcal P}$ specified in advance.
Under the fixed-precision formulation, by contrast, the sample
allocation adapts to the observed simulation outputs, creating
additional stochastic dependence.
The prespecified allocation allows us to characterize the effects of
means and correlations on $\mathrm{PCS}(\tau)$ directly.
Although the formal analysis focuses on the fixed-budget setting, its
qualitative insights also help explain the performance gains of P3C
under the fixed-precision formulation.

Before proceeding, we introduce additional notation. Let
$\sigma_i^2=\operatorname{Var}(X_i)$ denote the unknown variance of $X_i$
and $r_{ij}=\operatorname{Corr}(X_i,X_j)$ denote the unknown Pearson
correlation coefficient between $X_i$ and $X_j$. Throughout
the following analysis, we assume $r_{ij}\ge0$ for all
$i,j\in\mathcal P$. For any
$\tau\in\mathcal P$ and $i\in\mathcal P\setminus\{\tau\}$, define
\begin{equation}
\begin{aligned}
    \lambda_i^\tau
    &=\operatorname{Var}(\bar x_\tau-\bar x_i)
      =\frac{\sigma_\tau^2}{N_\tau}
       +\frac{\sigma_i^2}{N_i}
       -\frac{2\operatorname{Cov}(X_\tau,X_i)}{N_{\tau,i}},\\
    d_i^\tau
    &=\frac{\mu_\tau-\mu_i}{\sqrt{\lambda_i^\tau}},
    \qquad
    y_i^\tau
    =\frac{\bar x_\tau-\bar x_i-(\mu_\tau-\mu_i)}
           {\sqrt{\lambda_i^\tau}}.
\end{aligned}
\label{eq:local-standardized-differences}
\end{equation}
Then $\mathrm{PCS}(\tau)$ can be written as
$\Pr\{y_i^\tau>-d_i^\tau\text{ for all }
i\in\mathcal P\setminus\{\tau\}\}$.
The vector
$\bm y^\tau=(y_1^\tau,\ldots,y_{\tau-1}^\tau,
y_{\tau+1}^\tau,\ldots,y_p^\tau)$
follows $N(0,\Phi^\tau)$, where $\Phi^\tau$ is a
$(p-1)\times(p-1)$ covariance matrix with entries
$\tilde r_{i,j}^\tau=\operatorname{Corr}(y_i^\tau,y_j^\tau)$
for $i,j\in\mathcal P\setminus\{\tau\}$.
Its diagonal entries are one, and $|\tilde r_{i,j}^\tau|\leq1$.

We first present the technical results in
Subsection~\ref{sub_4.1.2}.
Subsection~\ref{sub_4.1.3} then provides an intuitive explanation of
the ``separation'' effect and shows how P3C uses correlation-based
clustering to improve selection efficiency. Finally, Subsection~\ref{sec4.2} quantifies the PCS improvement
induced by correlation-based clustering under a fixed sample
allocation. The proofs are provided
in the appendix.
}

\subsubsection{The derivative of PCS with respect to mean and correlation information.}\label{sub_4.1.2}

To begin with, the following fundamental property establishes that \(\mathrm{PCS}(\tau)\) prioritizes mean information and increases monotonically with respect to the mean \(\mu_\tau\) under any correlation configuration. 
\begin{theorem}\label{theo1}
 $\mathrm{PCS} \left(\tau\right)$
  is differentiable with respect to $\bm{\mu}=(\mu_1,...,\mu_p)$ and $\frac{\partial \mathrm{PCS}\left(\tau\right)}{{\partial\mu}_{\tau}}>0$.
\end{theorem}
Then we examine a more refined interaction between mean and correlation by analyzing the gradient with respect to correlation.
\begin{theorem}\label{theo2}
Let $\tau=[m]$. We define $\mathcal{I}^{+}(\tau)\triangleq\{[i]:i<m\}$ and $\mathcal{I}^{-}(\tau)\triangleq\{[i]:i>m\}$, which are the index sets of the alternatives with larger and smaller mean than $\tau$, respectively. We denote the sign of a real number as $\mathrm{sign}(\cdot)$. Then $\forall i\in\mathcal{P}\setminus\{\tau\}$, $$\frac{\partial PCS\left(\tau\right)}{\partial r_{\tau i}}=D_i^\tau+I_i^\tau,$$ 
where
 $$\ D_i^\tau\triangleq \frac{\partial \mathrm{PCS(\tau)}}{\partial d_i^\tau}\frac{\partial d_i{ }^{\tau}}{\partial r_{\tau i}},$$ 
  $$\ \  I_i^\tau=\sum_{j\in\{1,\cdots,p\}\setminus\{i,\tau\}} I_i^{\tau,j}\triangleq \sum_{j\in\{1,\cdots,p\}\setminus\{i,\tau\}} 2\frac{\partial \text{PCS}(\tau)}{\partial {{\widetilde{r}}_{i,j}}^{\tau} }\frac{\partial \widetilde{r}_{i,j}^{\tau}}{\partial r_{\tau i}}.$$

  Moreover, we have
  
   $\mathrm{(a)}$ $D_i^\tau>0$ for $i\in\mathcal{I}^-(\tau)$ and  $D_i^\tau<0$ for $i\in\mathcal{I}^+(\tau)$;
   
   $\mathrm{(b)}$ 
$\mathrm{sign}(I_i^{\tau,j}) = \mathrm{sign} \left( 
    -\sigma_i^2(N_i)^{-1} 
    + {r_{\tau,i} \sigma_\tau \sigma_i}{(N_{\tau,i})^{-1}} 
    - {r_{\tau,j} \sigma_\tau \sigma_j}{(N_{\tau,j})^{-1}} 
    + {r_{i,j} \sigma_i \sigma_j}{(N_{i,j})^{-1}} 
\right).$
  \end{theorem}

\begin{remark} The specific expressions of terms $D_i^\tau$ and $I_i^\tau$ can be found in the appendix. The analysis presented in this paper does not specify the computation of these terms but can help gain insights through the magnitude and sign of each term. To complete the analysis of the ``mean-covariance" interaction, we also present the impact of variance information in the appendix.  
\end{remark}
As shown in Theorem \ref{theo2}, the derivative $\frac{\partial PCS\left(\tau\right)}{\partial r_{i\tau}}$ is composed of two parts: the \textit{mean-determined} (MD) term $D_i^\tau$ and the \textit{mean-independent} (MI) term $I_i^\tau$. The sign of the $D_i^\tau$ depends on whether the mean $\mu_i$ is larger or smaller than $\mu_\tau$, whereas the sign of $I_i^\tau$ is independent of the mean. Determining the sign of $I_i^\tau$ is generally complex and unclear. Therefore, it is necessary to establish the magnitudes of $D_i^\tau$ and $I_i^\tau$ terms to ascertain which term dominates. For a given vector $\bm{d}$ and an index set $S$, we use $\bm{d}_S$ to denote the vector composed of the components from $\bm{d}$ with indices $S$. Similar notations $\Sigma_S$ are used for submatrices of matrix $\Sigma$. Let $N$ denote the total sample size, i.e., $N=\sum_{i\in\mathcal{P}}N_i$. 
\begin{assumption}\label{ass_infi}
   $N_i= \mathcal{O}\left(N_j\right)$ as $N\to\infty$, $\forall i,j\in\mathcal{P}$. 
 \end{assumption}

{\begin{remark}
Assumption~\ref{ass_infi} and Corollary~\ref{co1} concern the regime
$N\to\infty$ with $p$, the means, and the covariance matrix fixed.
This assumption is mild for fixed-budget R\&S procedures, as it
holds for any allocation policy that assigns every alternative a
positive limiting fraction of the total budget. For example,
\citet[Proposition~3.1]{wu2018analyzing} show that the allocation
proportions of classical OCBA and its variants OCBA-D and OCBA-R
converge almost surely to strictly positive limits under independent
normal sampling with distinct means and positive variances.
These algorithms therefore satisfy Assumption~\ref{ass_infi}
almost surely under those conditions.
\end{remark}
}

\begin{assumption} \label{ass_weak_cor}  $\forall i,j\in\mathcal{P}\setminus\{\tau\}$ such that $i\ne j$ and \( d_i^\tau \neq 0 \), the following conditions hold:
$\mathrm{(a)}$ $|\tilde{r}^\tau_{i,j}|< \big|\frac{d^\tau_j}{d^\tau_i}\big|$;
$\mathrm{(b)}$ $|\tilde{r}^\tau_{i,j}|< \big|\frac{d^\tau_j}{d^\tau_i}\big|^2$. 
 \end{assumption}

{
Assumption~\ref{ass_weak_cor} imposes a moderate-correlation condition on  
$\tilde r_{i,j}^\tau$. Here, $\tilde r_{i,j}^\tau$ denotes the correlation  
between the two pairwise comparisons $\bar x_\tau-\bar x_i$ and  
$\bar x_\tau-\bar x_j$, rather than the correlation between the 
performances of the alternatives themselves.
It characterizes the extent to which random fluctuations in one comparison 
(e.g., between alternatives $\tau$ and $i$) may affect the outcome of the 
other comparison (e.g., between alternatives $\tau$ and $j$).
If this correlation is excessively strong, a random fluctuation that causes 
an incorrect pairwise comparison may also induce a substantial error in the 
other comparison. The assumption rules out such extreme dependence between 
pairwise comparisons. 
 }

  \begin{corollary}
  \label{co1}
  
       $\forall i\in\mathcal{P}\setminus\{\tau\}$, we define $\mathcal{S}_{-i}^{+}(\tau)\triangleq\{j\in \mathcal{P}\setminus\{i,\tau\}|-\widetilde{d}_j^\tau>0\}$, $-\widetilde{d}_j^\tau\triangleq-d_j^\tau+d_i^\tau\widetilde{r}^\tau_{i,j}$, $j\in\mathcal{P}\setminus\{\tau,i\}$. If Assumption \ref{ass_infi} holds, as $N\to\infty$, then\\
      $\mathrm{(a)}$
      \begin{equation}\label{orderI}
        \left|D_i^\tau\right|= \mathcal{O}(e^{-\mathscr{D}^{\tau}_i N})\triangleq \mathcal{O}\bigg(\exp\big(-\frac{(d_i^\tau)^2+Q^\tau_S}{2}\big)\bigg),  
      \end{equation}
      \begin{equation}\label{orderE}
      \begin{aligned}
     \left|I_i^\tau\right|= \mathcal{O}(e^{-\mathscr{I}^{\tau}_i N})\triangleq \mathcal{O}\bigg( \exp{\big(-\min_{j\neq i}\frac{(d_i^\tau)^2+(d_j^\tau)^2}{2(1+|{{\widetilde{r}}_{i,j}}^{\tau}|)}\big)}\bigg),
      \end{aligned}
      \end{equation}
  where $$Q^\tau_S\triangleq\min_{\bm{x}\geq \bm{d}_{\mathcal{S}_{-i}^{+}(\tau)}}\langle \bm{x},(\Sigma^Z_{\mathcal{S}_{-i}^{+}(\tau)})^{-1}\bm{x} \rangle,$$ $$\bm{x}=(x_1,\cdots,x_{|\mathcal{S}_{-i}^{+}(\tau)|})\in\mathbb{R}^{|\mathcal{S}_{-i}^{+}(\tau)|},\ \ \bm{d}=(-\widetilde{d}_1^\tau,\cdots,-\widetilde{d}_p^\tau)\in\mathbb{R}^{p-2}$$   
  and $\Sigma^Z$ is a covariance matrix of $p$ transformed variables which are given in the appendix; \\
  $\mathrm{(b)}$ When $\tau=[1]$, if Assumption $\ref{ass_weak_cor}$$\mathrm{(a)}$ and $\ref{ass_weak_cor}$$\mathrm{(b)}$ hold, then $\mathscr{D}^{\tau}_i<\mathscr{I}^{\tau}_i$;\\
$\mathrm{(c)}$
Let \(\mathscr{D}^{\tau}_i(m)\) and \(\mathscr{I}^{\tau}_i(m)\) denote the asymptotic decay rates of $D_i^\tau$ and $I_i^\tau$ when $\tau=[m]$. Suppose only the ranking $m$ of $\tau$ changes, while all other parameters (covariance matrix, sample sizes, and the absolute value of pairwise mean differences) remain fixed. If Assumption $\ref{ass_weak_cor}$$\mathrm{(a)}$ holds,
\(\mathscr{D}^{\tau}_i(m)\) is non-decreasing as \(m\) increases, while \(\mathscr{I}^{\tau}_i(m)\) remains constant for all \(m\).
  \end{corollary}
Some key observations follow. First, both the MD and MI terms will decrease exponentially to 0 as the total sample size $N$ increases. Whether the MD terms or MI terms dominate depends on the relative magnitudes of decay rates $\mathscr{D}^{\tau}_i$ and $\mathscr{I}^{\tau}_i$. Second,
combining Corollary \ref{co1}(b) and Corollary \ref{co1}(c), let $\tau=[m]$. When \( m = 1 \), i.e., {when \(\text{PCS}(\tau) = \text{PCS}_{\text{trad}}\), the MD term is dominant}. As the ranking $m$ increases, the MI term gradually surpasses the MD term. While the exact ranking $m$ at which the MI term becomes the dominant component is unclear, we can at least confirm that there exists a subset $\mathcal{P}'\subseteq \mathcal{P}$ containing $[1]$, within which the MD term remains dominant for all alternatives. The alternatives in $\mathcal{P}'$ are top-ranked ``good" alternatives, and in practice, users are typically only interested in these ``good" ones. Therefore, in the intuitive analysis in Section \ref{sub_4.1.3}, we assume that the MD term dominates for any alternative under consideration (this assumption is used exclusively in the intuitive analysis below).

\subsubsection{``Separation" effect}\label{sub_4.1.3}
  \begin{figure}
{
\centering
\includegraphics[width=0.8\textwidth]{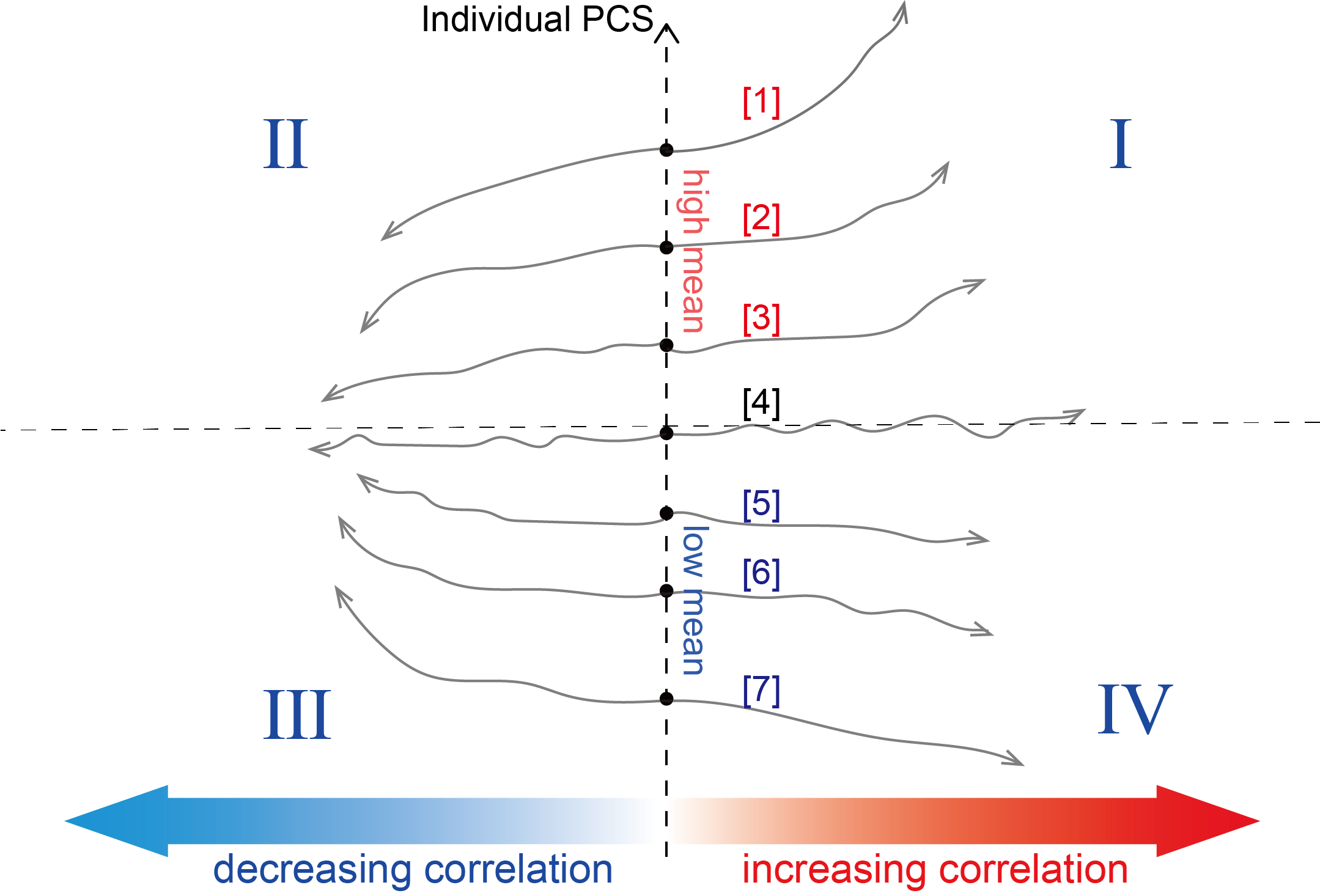}
\caption{An illustrative example with 7 alternatives, where the top 3 alternatives $[1]$, $[2]$ and $[3]$ exhibit high mean performances, and $[5]$, $[6]$ and $[7]$ exhibit low mean performances.\label{figtheory}}
}
\end{figure}

Based on the aforementioned technical results, we are now ready to show the underlying insight.
According to Theorem \ref{theo2}(a), focusing solely on the MD term, the correlations between $\tau$ and alternatives in $\mathcal{I}^-(\tau)$ impose a cumulative positive effect on $\text{PCS}\left(\tau\right)$, and conversely, correlations between $\tau$ and those alternatives in $\mathcal{I}^+(\tau)$ impose a cumulative negative effect. Consequently, the impact of correlation on $\text{PCS}\left(\tau\right)$ hinges on the ranking of $\tau$. If $\tau$ is near the top, the number of alternatives in $\mathcal{I}^-(\tau)$ dominates, and then the cumulative positive effects surpass the cumulative negative effects. Increasing the correlations $\{r_{\tau i}\}_{i\in\mathcal{P}\setminus\{\tau\}}$ is more likely to improve $\text{PCS}(\tau)$ (specifically, when $\tau=[1]$ and $\mathcal{I}^+(\tau)=\emptyset$, then increasing the correlation will certainly improve $\text{PCS}_{\text{trad}}$). On the contrary, if the alternative $\tau$’s mean performance is poor, the influence of correlation is reversed, and increasing the correlation around $\tau$ will decrease $\text{PCS}(\tau)$. 

In summary, as shown in regions \uppercase\expandafter{\romannumeral1} and \uppercase\expandafter{\romannumeral4} 
 of Figure \ref{figtheory}, increasing the correlation level among the alternatives can induce a ``separation" effect, which may ``amplify" good alternatives and ``suppress" bad alternatives in terms of individual PCS.
 This makes the distinction between good and bad alternatives more pronounced, thereby accelerating the R\&S procedure. Conversely, decreasing the overall correlation leads to an undesirable aggregation phenomenon (regions \uppercase\expandafter{\romannumeral2} and \uppercase\expandafter{\romannumeral3}). In practice, we cannot directly modify the correlation parameters within the true distribution of alternatives. However, in a parallel computing environment, we can enhance the local correlation level in each processor using clustering techniques. This is the rationale behind P3C's use of correlation-based clustering to enhance sample efficiency.

Before concluding this section, we present the following lemma, which will be used in the next section as a supplement to Theorem \ref{theo2}. 
\begin{lemma} \label{lemma1}
   $\forall i,j\in\mathcal{P}\setminus\{\tau\}$, $\frac{\partial \text{PCS}\left(\tau\right)}{\partial r_{ij}}\geq 0$. If $\mathscr{D}^{\tau}_i<\mathscr{I}^{\tau}_i$, then $$\frac{\partial \text{PCS}\left(\tau\right)}{\partial r_{ij}}= o\bigg(|\frac{\partial \text{PCS}\left(\tau\right)}{\partial r_{\tau i}}|\bigg).$$
\end{lemma}
This lemma characterizes the impact of correlations $\{r_{ij}\}_{i,j\in\mathcal{P}\setminus\{\tau\}}$, which are \textit{not} directly associated with $\tau$. These correlations in the surrounding environment always have a non-negative impact on $\text{PCS}(\tau)$, regardless of mean information. This further complements the argument that clustering correlated alternatives can accelerate the R\&S procedure. Additionally, if $\tau$ is a ``good" alternative within $\mathcal{P}'$, meaning that $\mathscr{D}^{\tau}_i<\mathscr{I}^{\tau}_i$, the impact of $\{r_{ij}\}_{i,j\in\mathcal{P}\setminus\{\tau\}}$ is negligible compared to  $\{r_{\tau i}\}_{i\in\mathcal{P}\setminus\{\tau\}}$, so it will not affect the aforementioned ``separation" phenomenon induced by $\{r_{\tau i}\}_{i\in\mathcal{P}\setminus\{\tau\}}$.

{  

\subsubsection{Quantifying PCS Improvement under a Fixed Sample Allocation}
\label{sec4.2}

After characterizing the mean--correlation interaction, we now quantify the PCS improvement induced by correlation-based clustering under a fixed-budget setting. We consider a stylized symmetric benchmark with a prescribed deterministic sample allocation, formalized in the following assumption.

\begin{assumption}[Stylized symmetric benchmark]
\label{ass_symmetric_benchmark}
\begin{enumerate}
    \item[(a)] The $p$ alternatives are partitioned into $k$ equally sized clusters
    $\mathcal G_1,\ldots,\mathcal G_k$, where $k\geq2$ and $p/k$ is an integer greater than one. Each cluster $\mathcal G_j$ contains one local best alternative $\tau_j$ and $p/k-1$ suboptimal alternatives.

    \item[(b)] The local best alternatives share the same mean and variance, and the suboptimal alternatives also share the same mean and variance. The common mean of the local best alternatives is strictly greater than that of the suboptimal alternatives.

    \item[(c)] The correlation coefficient is $R$ between alternatives in the same cluster and $r$ between alternatives in different clusters, where
    $\Delta r=R-r>0$.

    \item[(d)] The true cluster partition is unknown, but the indices of the local best alternatives are known. Each local best $\tau_j$ is fixed on processor $j$.

    \item[(e)] There are $k$ processors, and each correlation-based cluster is assigned to a separate processor. Each processor is assigned exactly $p/k-1$ suboptimal alternatives in addition to its fixed local best. The sample allocation is deterministic and identical across processors, with a prescribed sample size for each local best and a common prescribed sample size for each suboptimal alternative. The total budget available for selection is $N$; it excludes the
samples used for clustering.
\end{enumerate}
\end{assumption}

Except for the $k$ local best alternatives, 
we compare two strategies for assigning the remaining $p-k$ sub-optimal alternatives to the processors. 
Strategy $\mathcal R$ assigns them uniformly at random, whereas strategy $\mathcal C$ assigns alternatives from the same correlation-based clusters to the same processor.
Unlike the general problem formulation, this benchmark does not assume a unique global best among all $p$ alternatives because $\tau_1,\ldots,\tau_k$ have the same mean. Nevertheless, each $\tau_j$ has a strictly larger mean than every suboptimal alternative and is therefore the unique best in the local selection problem on processor $j$. For analytical convenience, Assumption~\ref{ass_symmetric_benchmark} sets the number of clusters $k$ equal to the number of available processors $k_m$. In practice, $k$ and $k_m$ may differ. When $k>k_m$, a standard master--worker framework can dynamically assign multiple clusters to each processor, as in existing parallel R\&S implementations \citep{luo2015fully}.

Since only finite samples are used to learn the correlation structure, 
the resulting clustering may be imperfect. Let $G(i)$ and $G_n(i)$ denote the true and estimated cluster labels of
alternative $i$, respectively, where $G_n(i)$ is the output of the
clustering algorithm based on $n$ samples per alternative. Writing 
$\Pi=(G(1),\ldots,G(p))$ and 
$\Pi_n=(G_n(1),\ldots,G_n(p))$, 
we define the probability of correct clustering (PCC) as
$$
\mathrm{PCC}\triangleq\mathbb{P}(\Pi_n=\Pi).
$$
We suppress its dependence on $n$ in the notation.

We focus on the PCS of the first processor. Let $\mathrm{PCS}_{\mathcal R}^{1}$ and $\mathrm{PCS}_{\mathcal C}^{1}$
denote the probabilities of selecting $\tau_1$ on the first processor
under strategies $\mathcal R$ and $\mathcal C$, respectively.
Under strategy $\mathcal C$, after applying the correlation-based
clustering, let $\widehat Q_n$ denote the proportion of suboptimal
alternatives assigned to the first processor that belong to the same
true cluster as the local best $\tau_1$:
$$
\widehat Q_n
=
\frac{
\sum_{i\in\mathcal G_1\setminus\{\tau_1\}}
\mathbf 1\{G_n(i)=1\}
}{
p/k-1
}.
$$
If $\mathbb E[\widehat Q_n]>1/k$, moving from Strategy~$\mathcal R$
to Strategy~$\mathcal C$ places, on average, more alternatives from
$\mathcal G_1$ on the first processor. This is equivalent to increasing the average
correlation between $\tau_1$ and its local competitors. Through the
mean--correlation interaction established above, it improves the
probability of selecting $\tau_1$. Theorem \ref{theo_saving} quantifies this improvement.

\begin{theorem}[PCS improvement]
\label{theo_saving}Suppose that Assumptions~\ref{ass_infi} and
\ref{ass_symmetric_benchmark} hold.
Suppose also that Assumption~\ref{ass_weak_cor}(a) and the condition
$\mathscr D_i^{\tau_1}<\mathscr I_i^{\tau_1}$ in
Lemma~\ref{lemma1} hold. Suppose that
$\mathbb E[\widehat Q_n]>1/k$.
For sufficiently large $N$, the expected PCS
improvement satisfies
\begin{equation}
\label{complex_reduc}
\mathbb E\!\left[
    \mathrm{PCS}_{\mathcal C}^{1}
    -
    \mathrm{PCS}_{\mathcal R}^{1}
\right]
\geq
\gamma_N\Delta r
\left(\frac{p-k}{k}\right)
\left(
    \mathbb E[\widehat Q_n]-\frac1k
\right)>0,
\end{equation}
where $\gamma_N>0$ is defined in the appendix.
\end{theorem}
According to Theorem~\ref{theo_saving}, correlation-based clustering
improves PCS whenever
$\mathbb E[\widehat Q_n]>1/k$.
This condition means that the expected fraction of suboptimal
alternatives assigned to the first processor that belong to the same
true cluster as $\tau_1$ exceeds the corresponding fraction under
random assignment.
In other words, PCS improvement can be achieved as long as the
clustering step performs better than random assignment in this sense.
This condition can also be related to PCC.
By definition, $\Pi_n=\Pi$ implies $\widehat Q_n=1$, and hence
$
\mathrm{PCC}
\leq
\mathbb E[\widehat Q_n].
$
Since PCC requires the cluster labels of all alternatives to be
recovered correctly, it is a more stringent measure of clustering
quality, and $\mathbb E[\widehat Q_n]$ can be substantially larger
than PCC.
Consequently, $\mathrm{PCC}>1/k$ is sufficient, but not necessary,
for a positive PCS improvement.

Although Theorem~\ref{theo_saving} is established under the
fixed-budget formulation with a prespecified sample
allocation, the mean--correlation interaction underlying this result
has a natural implication for the fixed-precision formulation:
grouping highly correlated alternatives may reduce the expected
sample size required to attain a prescribed PCS.
We investigate this implication empirically in
Section~\ref{chap7.2}.
}

\section{Large-Scale Alternative Clustering}
\label{chap6}

{  Although PCC does not determine the sample optimality of P3C,
Theorem~\ref{theo_saving} shows that better correlation-based
clustering can lead to a larger PCS improvement. }Nevertheless, clustering in large-scale problems presents challenges, both computationally and statistically. To address these challenges and strive for a high PCC, inspired by a well-known few-shot learning approach in the deep learning literature, namely, Prototypical Networks \citep{snell2017prototypical}, we propose a parallel alternative clustering algorithm $\mathcal{A}\mathcal{C}^+$. Following this, we detail the methodology for calculating the PCC.
 
\subsection{Few-shot Alternative Clustering}
\label{chap6.2}
In P3C, we cluster alternatives instead of their simulation outputs, a concept known as variable clustering in statistical literature. A commonly used algorithm for this purpose is the hierarchical clustering algorithm \citep{jolliffe1972discarding}, denoted as $\mathcal{A}\mathcal{C}$  (see pseudocode in the appendix). The framework of $\mathcal{A}\mathcal{C}$ is as follows. Suppose the number of clusters $k$ is known and each alternative has a sample size of \( n \). Initially, each alternative is treated as an individual group. In each iteration, the two groups with the maximum similarity $R$ are merged until $k$ groups remain.  
The (empirical) similarity between two groups of alternatives, $G^1$ and $G^2$, is quantified by $R\left(G^1,G^2\right)=\max \limits_{i\in G^1,j\in G^2}\hat{r}_{ij}^n$, where $\hat{r}_{ij}^n$ denotes the estimated $r_{ij}$ using $n$ samples.
However, $\mathcal{A}\mathcal{C}$ still requires estimating the entire $p\times p$ correlation matrix, which necessitates a large sample size, and more importantly, cannot be parallelized.

\begin{algorithm}[htbp]
\caption{Few-shot Alternative Clustering $\mathcal{AC}^+$}
\label{acplus}
\begin{algorithmic}[1]

\State \textbf{Input:} sample size $n$, number of clusters $k$, $p_s$, $p_q$.

\State \textbf{Step 1 (Splitting):}
Randomly split the alternative set $\mathcal{P}$ into two subsets
$\mathcal{P}_s$ and $\mathcal{P}_q$ with sizes $p_s$ and $p_q$, respectively.

\State {  \textbf{Step 2 (Selecting Prototypes):}
On a single processor, simulate $n$ samples for each alternative in
$\mathcal{P}_s$ and compute the estimated covariance matrix
$\hat{\Sigma}_n^{\mathcal{P}_s}$.
Apply the clustering algorithm $\mathcal{AC}$ to
$\hat{\Sigma}_n^{\mathcal{P}_s}$ and obtain the clustering result
$\Pi_n^{\mathcal{P}_s}$.
For each cluster $\mathcal{G}_j$, $j\in\{1,2,\dots,k\}$, compute its
first principal component
$PC_{\mathcal{G}_j}
=\sum_{i\in\mathcal{G}_j}k_{ij}X_i$,
where $k_{ij}$ denotes the loading of alternative $i$ on
$PC_{\mathcal{G}_j}$.
Select the prototype as
$\tau_j=\arg\max_{i\in\mathcal{G}_j}k_{ij}$.
Let $\mathcal{P}_p=\{\tau_1,\dots,\tau_k\}$.}

\State \textbf{Step 3 (Matching):}
Send a copy of $\mathcal{P}_p$ to each processor.
Randomly and evenly allocate $\mathcal{P}_q$ across processors.
For each $j \in \mathcal{P}_q$, simulate $n$ samples of alternative $j$
and compute $\hat r_{\tau_i,j}^n$ for all $i \in \{1,2,\dots,k\}$.
Assign
\[
G_n(j) = \arg\max_{i \in \{1,2,\dots,k\}} \hat r_{\tau_i,j}^n .
\]

\State \textbf{Output:} Return the partition $\Pi_n$.

\end{algorithmic}
\end{algorithm}
  
The few-shot $\mathcal{A}\mathcal{C}^+$ algorithm resolves the above issues. As shown in Algorithm \ref{acplus}, initially, $\mathcal{P}$ is split into two sets: the support set $\mathcal{P}_s$ and the query set $\mathcal{P}_q$, with sizes $p_s$ and $p_q$ respectively. $p_s$ is set to be greater than $k$ but much smaller than $p$. Next, alternatives in $\mathcal{P}_s$ are grouped into $k$ clusters using the $\mathcal{A}\mathcal{C}$ algorithm, which can be efficiently handled on a single processor due to the moderate size of $p_s$. In each cluster $\mathcal{G}_j$ ($j=1,2,\cdots,k$), one \textit{representative} alternative $\tau_j$ is chosen as the ``prototype" of the cluster. Finally, each alternative in the query set $\mathcal{P}_q$ is assigned to an existing cluster by identifying the most correlated prototype. This matching process can be parallelized by sending a copy of the $k$ \textit{prototype}s to each processor and then randomly assigning alternatives in $\mathcal{P}_q$ to different processors. {  Given that $k$ is typically much smaller than $p$, the additional simulation cost for these copies is negligible. $\mathcal{A}\mathcal{C}^+$ eliminates the need for estimating the entire correlation matrix and only requires estimating a small submatrix.}

As for the selection of the ``prototype" within each cluster $\mathcal{G}_j$, the process is as follows. (\romannumeral1) Apply principal component analysis (PCA) to the cluster $\mathcal{G}_j$ to identify the first principal component $PC_{\mathcal{G}_j}$, which is a synthetic variable given by $PC_{\mathcal{G}_j}=\Sigma_{i\in\mathcal{G}_j}k_{ij}X_{i}$. Detailed calculations can be found in standard machine learning textbooks. In variable clustering literature, $PC_{\mathcal{G}_j}$ is often termed as the ``latent component" of $\mathcal{G}_j$, proven to be the linear combination that maximizes the sum of squared correlations with the alternatives located in $\mathcal{G}_j$ \citep{vigneau2003clustering}. The loading $k_{ij}$ measures the correlation between the alternative $i$ and $PC_{\mathcal{G}_j}$. (\romannumeral2) Select the alternative with the largest loading $k_{ij}$ as the \textit{representative} ``prototype" of $\mathcal{G}_j$ \citep{al2001variable}: 
 $ \tau_j= \mathop{\arg\max}_{i\in \mathcal{G}_j}k_{ij}.$

Moreover, as shown in Proposition \ref{prop:fp-sample-optimality}, bounding the number of alternatives within each cluster by \( p_m \) is essential for asymptotic sample optimality. Fortunately, the hierarchical structure of the \(\mathcal{A}\mathcal{C}\) and \(\mathcal{A}\mathcal{C}^+\) algorithms makes this easy to enforce. We set an upper limit \( p_m \) on cluster size—once reached, additional alternatives are assigned to the second closest cluster.  
{  The clustering algorithms described above require the number of clusters $k$ to be specified in advance. The upper limit $p_m$ on the number of alternatives in each cluster provides practical guidance for choosing $k$: it can be set to approximately 2--3 times $\frac{p}{p_m}$.}
 In practice, the choice of \( k \) is flexible, as the exact number of clusters is not critical so long as highly correlated alternatives are grouped together. In addition, in the appendix, we describe two additional steps to address potential issues: (i) unbalanced cluster sizes and (ii) the presence of alternatives with similar means within the same cluster.

\subsection{Computation of $\mathrm{PCC}$}\label{secpcc}
Next, we establish a computable lower bound for $\mathrm{PCC}_{\mathcal{A}\mathcal{C}^+}$, the statistical guarantee of $\mathcal{A}\mathcal{C}^+$. 
 {First, we assume that, alternatives within the same cluster are more highly correlated than those across clusters, i.e., the correlation structure satisfies the following Assumption~\ref{ass1}.
\begin{assumption}\label{ass1} 
 $r_{ab}>r_{cd}$ for all $a, b, c, d \in \mathcal{P}$ such that $G(a) = G(b)$ and $G(c) \neq G(d)$.
\end{assumption}}

Let $\Gamma\triangleq\{(ab,ac)|G(a)=G(b), G(a)\neq G(c), a,b,c \in\mathcal{P}, a\neq b\}$ be the collection of pairs of overlapping intra-cluster and inter-cluster correlations. $\Gamma_{s}\triangleq \{(ab,ac)\in\Gamma|a,b,c\in\mathcal{P}_s\}$ and $\Gamma_{q}\triangleq\{(a\tau_{i},a\tau_{j})\in\Gamma| a\in \mathcal{P}_q, i=G(a), j\in \{1,2,\dots k\}\setminus \{i\}\}$ are two subsets of $\Gamma$. Fisher's $z$ transformation of correlation coefficients, defined as $$z\left(r\right)\triangleq\frac{1}{2}ln\frac{1+r}{1-r},$$ is monotonically increasing in $(0,1)$. Therefore, Assumption \ref{ass1} implies that $z(r_{ab})>z(r_{ac})$ for any pair $(ab,ac)\in \Gamma$. Borrowing the idea of indifference zone in R\&S literature, in the following proposition we introduce a correlation indifference parameter $\delta_c>0$ and assume $z(r_{ab})>z(r_{ac})+\delta_c$.

\begin{proposition}\label{theo_clust}
{If Assumption \ref{ass1} holds, all clusters have equal sizes, the number of clusters $k$ is known,} and the $\mathcal{A}\mathcal{C}^+$ algorithm uses $n$ samples to perform clustering, then 
 \begin{equation}
\label{acpluspcc}
   \mathrm{PCC}_{\mathcal{A}\mathcal{C}^+}\geq \big(1-k(1-1/k)^{p_s}\big)P\bigg(\bigcap_{(ab,ac)\in\Gamma_{s}} \{\hat{r}_{ab}^n >\hat{r}_{ac}^n\}\bigg)P\bigg(\bigcap_{(ab,ac)\in\Gamma_{q}} \{\hat{r}_{ab}^n >\hat{r}_{ac}^n\}\bigg). 
\end{equation}
Let $\Gamma_{\star}$ be either $\Gamma_s$ or $\Gamma_q$. If $z(r_{ab})>z(r_{ac})+\delta_c$ for any $(ab,ac)\in\Gamma_{\star}$, then
 \begin{equation}
\label{acpluspcc2}
P\bigg(\bigcap_{(ab,ac)\in\Gamma_{\star}} \{\hat{r}_{ab}^n >\hat{r}_{ac}^n\}\bigg)\geq\sum_{(ab,ac)\in\Gamma_{\star}}{\Phi\Bigg({\delta_c}{\sqrt{\frac{n-3}{2(1-\hat{r}_{bc}^n)h(a,b,c)}}}\Bigg)}-(\left\lvert \Gamma_{\star}\right\rvert-1), 
\end{equation}
where $h(a,b,c)=\frac{1-f\cdot\bar{R}^2}{1-{\bar{R}}^2}$, with $\bar{R}^2(a,b,c)=\frac{\left(\hat{r}_{ab}^n\right)^2+\left(\hat{r}_{bc}^n\right)^2}{2}$ and $f(a,b,c)=\frac{1-\hat{r}_{bc}^n}{2\left(1-\bar{R}^2\right)}$.  
\end{proposition}

Additionally, the computation of $\mathrm{PCC}_{\mathcal{A}\mathcal{C}^+}$ when cluster sizes are \textit{unequal} can be found in the appendix. {As $n\to\infty$ (with $p$ fixed), the lower bound of $\mathrm{PCC}_{\mathcal{A}\mathcal{C}^+}$ converges to $1-k(1-\frac{1}{k})^{p_s}$. This bound approaches 1 as \( p_s \) increases.
 P3C enjoys the advantage of not requiring an excessively high PCC or accurate correlation estimation. As long as the accuracy surpasses that of random clustering, an improvement in sample efficiency is ensured.
 Proposition \ref{theo_clust} provides a theoretical guarantee on the clustering 
quality when the number of clusters $k$ is available or can be reasonably 
specified. }

\section{Numerical Experiments}\label{sec_exp}
In Section \ref{chap7.1}, we present a simple example to illustrate the key theoretical result of this paper: the mean-correlation interaction (Theorem \ref{theo2}, Corollary \ref{co1}, and Lemma \ref{lemma1}). Section \ref{chap7.2} evaluates P3C under fixed-precision constraints, while Section \ref{chap7.3} examines its performance under fixed-budget constraints. {Throughout Sections~\ref{chap7.2} and~\ref{chap7.3}, all reported total
sample sizes and simulation budgets include the initialization samples
used to estimate correlations and perform clustering.}
\subsection{Illustrative Example}
\label{chap7.1}
Consider a group of 5 alternatives: $(X_1,X_2,X_3,X_4,X_5)$, each following a normal distribution. The mean and covariance parameters are given by
$\bm{\mu}=(\mu_1,\mu_2,\mu_3,\mu_4,\mu_5)=(2.1,2.0,1.95,1.9,1.9)$ and 
\[
\renewcommand{\arraystretch}{1.5} %
\Sigma =
\left(\begin{array}{ccccc}
0.1 & x & x & x & x \\
x & 0.1 & 0.01 & 0.01 & y \\
x & 0.01 & 0.1 & 0.01 & y \\
x & 0.01 & 0.01 & 0.1 & y \\
x & y & y & y & 0.1
\end{array}\right).
\]
All alternatives share the same variance. Alternative 1 has the highest mean, and its covariance with other alternatives is $x$ ($0\leq x<0.1$). Alternative 5 is among the worst performers, and its covariance with alternatives 2, 3 and 4 is $y$ ($0\leq y<0.1$). To illustrate the proposed mean-correlation interaction theory (Theorem \ref{theo2}, Corollary \ref{co1} and Lemma \ref{lemma1}), we calculate $\text{PCS}(1)$ and  $\text{PCS}(5)$ with different $x$ and $y$, keeping the mean and variance parameters constant. The sample size for each alternative is 10 (total sample size $N=50$). Table \ref{tab_meancov} presents the values of $\text{PCS}(1)$ and $\text{PCS}(5)$ when $x$ takes on the values of 0.01, 0.02, 0.03, and 0.05, and $y$ takes on the values of 0, 0.02, 0.04, and 0.06, respectively (there is no value in the table for ``$x=0.05, y=0.06$" because $\Sigma$ is not positive definite). The actual values of individual PCS are calculated using Monte Carlo numerical integration. The experimental results align with our theory of mean-correlation interaction: increasing correlation promotes alternatives with high means while suppressing those with low means. When $y$ is held constant, $\text{PCS}(1)$ increases as $x$ increases, consistent with Theorem \ref{theo2} and Corollary \ref{co1}. When $x$ is fixed, $\text{PCS}(1)$ also increases with $y$, where $y$ represents correlations between alternative 5 and alternatives 2-4, not directly associated with alternative 1. This result is consistent with Lemma \ref{lemma1}. Conversely, $\text{PCS}(5)$ decreases as $x$ (or $y$) increases, as predicted by Theorem \ref{theo2}.
  
\begin{table}[h]
  \centering
  \caption{$\text{PCS}(1)$ and $\text{PCS}(5)$ with Different $x$ and $y$ ($N=50$).}
  \renewcommand{\arraystretch}{2.5} 
  \setlength{\tabcolsep}{4pt} 
  \small 
  \begin{tabular}{c|ll|ll|ll|ll} 
    \midrule\midrule
    \multirow{2}{*}{}  & \multicolumn{2}{c|}{$y=0$} & \multicolumn{2}{c|}{$y=0.02$} & \multicolumn{2}{c|}{$y=0.04$} & \multicolumn{2}{c}{$y=0.06$} \\ \cline{2-9} 
    & \multicolumn{1}{c}{$\text{PCS}(1)$} & \multicolumn{1}{c|}{$\text{PCS}(5)$} 
    & \multicolumn{1}{c}{$\text{PCS}(1)$} & \multicolumn{1}{c|}{$\text{PCS}(5)$} 
    & \multicolumn{1}{c}{$\text{PCS}(1)$} & \multicolumn{1}{c|}{$\text{PCS}(5)$} 
    & \multicolumn{1}{c}{$\text{PCS}(1)$} & \multicolumn{1}{c}{$\text{PCS}(5)$} \\
    \cmidrule{1-9} 
    $x=0.01$ & 0.6707 & 0.0349 & 0.6741 & 0.0269 & 0.6875 & 0.1777 & 0.6879 & 0.0051 \\\cline{2-9}
    $x=0.02$ & 0.6852 & 0.0331 & 0.6904 & 0.0255 & 0.7003 & 0.0168 & 0.7018 & 0.0050 \\\cline{2-9}
    $x=0.03$ & 0.6979 & 0.0288 & 0.7141 & 0.0233 & 0.7171 & 0.0154 & 0.7274 & 0.0048 \\\cline{2-9}
    $x=0.05$ & 0.7506 & 0.0197 & 0.7662 & 0.0166 & 0.7690 & 0.0119 & \multicolumn{2}{c}{\diagbox[width=4.8em,height=0.6em]{}} \\ 
    \midrule\midrule
  \end{tabular}
  \label{tab_meancov}
\end{table}

\subsection{Fixed-precision R\&S: Drug Discovery}
\label{chap7.2}
We test the performance of P3C on the narcotic analgesics drug discovery problem introduced in \cite{negoescu2011knowledge}, which studies a set of $11\times8 \times 5 \times 6 \times 11 \times 3 \approx 10^5$ Benzomorphans, each representing a potential alternative drug. Simulation data are generated based on the Free-Wilson model \citep{free1964mathematical}: each drug’s value is modeled as the sum of the values of its constituent atomic groups. The value of each atomic group is treated as an independent normally distributed random variable, with mean estimated via regression on experimental data from \cite{katz1977application}, and variance randomly drawn from \(N(0, 0.1)\), excluding negative values. Since the Free-Wilson model inherently uses CRN, structurally similar drugs tend to be highly correlated. 
We adopt the fixed-precision formulation of R\&S. Our objective is to identify the best drug while ensuring $\mathrm{PCS}_{\mathrm{trad}}>0.9$. 
In this experiment, we compare P3C-KN and P3C-KT with the traditional stage-wise procedure from \cite{rinott1978two}, the Good Selection Procedure (GSP) in \cite{ni2017efficient}, and the standard KT and KN without P3C. The indifference-zone parameter is set to \(\delta = 0.1\), with \(p_m = 1000\) and \(k = 104\).
The initialization sample size is set to \(N_0 = 20\), and both correlation estimation and clustering in P3C are performed solely using these initialization samples, without requiring any additional sampling.
Experiments are conducted on a commercial cloud platform using a computing cluster with 104 processors. We adopt MATLAB’s \textit{Parallel Computing Toolbox} to manage workload distribution and inter-processor communication. For fairness, Rinott, GSP, standard KT, and standard KN are all implemented in the same parallel environment as P3C, using the traditional divide and conquer strategy and CRN. Each experiment is repeated 20 times, and results are averaged.

\begin{figure}[t]
    \centering
    \includegraphics[width=0.7\textwidth]{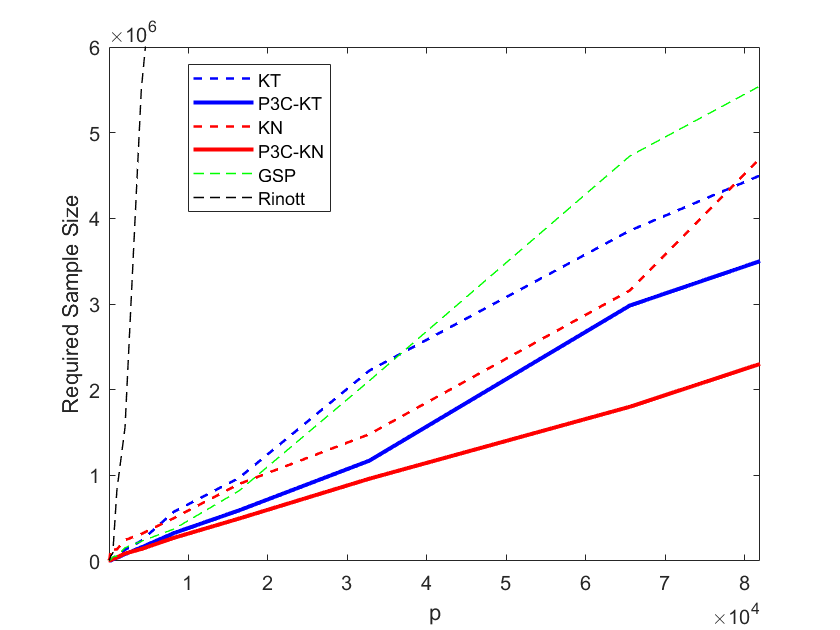}
    \caption{
    Required total sample size as a function of the number of alternatives \(p\).
    The vertical axis reports the total number of samples, and the horizontal axis is \(p\).
    }
    \label{fig:drug_sample}
\end{figure}

{Figure~\ref{fig:drug_sample} reports the total sample sizes required by P3C-KT, P3C-KN, KT, KN, Rinott, and GSP as \(p\) increases. P3C-KN and P3C-KT consistently outperform the other algorithms. Both algorithms exhibit linear growth in sample size with respect to \(p\), confirming the \(\mathcal{O}(p)\) sample complexity predicted by Proposition~\ref{prop:fp-sample-optimality}.
On average, P3C-KN requires only 46\% of the sample size used by
standard KN, while P3C-KT uses 68\% of that required by KT.
These results support the analysis in Section~\ref{chap4}, which
explains how correlation-based clustering in P3C can improve sample
efficiency in parallel R\&S. Notably, the slope of KT decreases visibly after applying P3C,
suggesting that, although KT already achieves the sample-optimal
$O(p)$ order, P3C can further reduce its leading constant and improve
finite-sample performance.}
In contrast, the classical stage-wise Rinott procedure demands 30–50 times more samples than the other fully sequential methods. Table~\ref{tab:drug_time} reports the wall-clock time of the experiments. Compared to the standard ``divide and conquer" framework, P3C introduces only one additional clustering step, making clustering the primary source of extra overhead. The table includes clustering time using the \(\mathcal{AC}^+\) algorithm, where \(p_s\) is set to one-tenth of the total alternatives. Even when $p$ reaches $2^{16}$, clustering takes only 11.714 seconds, which is negligible relative to the total R\&S runtime.
Notably, P3C-KT and P3C-KN are significantly faster than their non-P3C counterparts. This is because the reduction in sample complexity achieved by P3C far outweighs the clustering cost.

\begin{table}[t]
    \centering
    \caption{
    Wall-clock running time (in seconds) for different procedures under varying numbers of alternatives \(p\).
    }
    \label{tab:drug_time}
    \renewcommand{\arraystretch}{1.05}   
    \setlength{\tabcolsep}{5pt}          
    \begin{tabular}{c|cccc}
        \midrule\midrule
        \diagbox[width=9em,height=1.2em]{}{$p$} 
        & $2^{11}$ & $2^{13}$ & $2^{15}$ & $2^{16}$ \\ 
        \midrule
        \textbf{Clustering Time} & \textbf{0.086} & \textbf{0.208} & \textbf{2.438} & \textbf{11.714} \\ 
        \textbf{KT}      & 2.213 & 9.379  & 47.708 & 129.790 \\
        \textbf{P3C-KT}  & 1.288 & 3.306  & 18.188 & 48.830 \\
        \textbf{KN}      & 3.708 & 12.057 & 65.770 & 168.114 \\
        \textbf{P3C-KN}  & 8.123 & 17.803 & 37.573 & 89.200 \\ 
        \textbf{GSP}     & 2.646 & 6.686  & 71.556 & 278.306 \\
        \midrule\midrule
    \end{tabular}
\end{table}

\begin{figure}[htbp]
\centering
\includegraphics[width=0.7\textwidth]{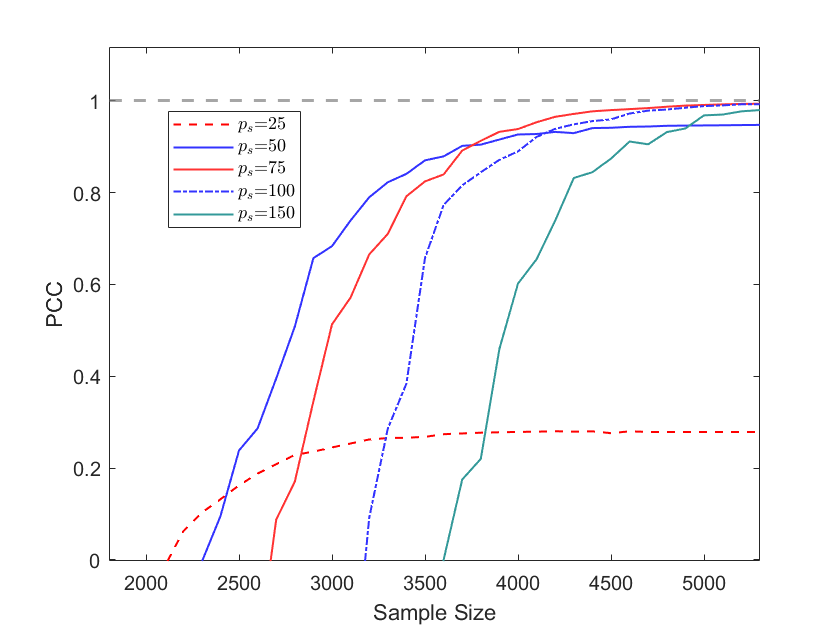}
\caption{Comparison of \( \mathrm{PCC}_{\mathcal{A}\mathcal{C}^+} \) with different support set sizes \( p_s \).}
\label{fig_pccexp}
\end{figure}

{
Theorem~\ref{theo_saving} shows that the PCS improvement achieved by
P3C depends on the quality of correlation-based clustering.
We use $\mathrm{PCC}$ as a metric of clustering quality in the
following experiments.} Using the $\mathcal{A}\mathcal{C}^+$ algorithm, we investigate $\mathrm{PCC}_{\mathcal{A}\mathcal{C}^+}$ under different sample sizes and different sizes of the support set ($p_s=25, 50, 75, 100, 150$). We focus on the first 1024 drugs and set the cluster number to $k=8$. The results are shown in Figure \ref{fig_pccexp}. As the sample size approaches infinity, consistent with Proposition \ref{theo_clust}, $\mathrm{PCC}_{\mathcal{A}\mathcal{C}^+}$ converges to $1-k(1-1/k)^{p_s}$ (denoted as $P(D)$), which is strictly less than 1. Larger $p_s$ values lead to a larger $P(D)$. When $p_s=25$, the PCC is notably constrained by the $P(D)$ term and cannot exceed 0.3. As $p_s$ increases to 50, $\mathrm{PCC}$ approaches acceptably close to 1.

\subsection{Fixed-budget R\&S: Neural Architecture Search}
\label{chap7.3}

Neural Architecture Search (NAS), a key challenge in deep learning, aims to identify the best-performing neural network architecture for a specific task, given that evaluating a neural architecture on a test dataset is computationally expensive.
The general NAS problem can be divided into two phases: architecture search (including the search space and search strategy) and architecture performance evaluation.
Suppose that we have already obtained a set $\mathcal{P}$ of $p$ alternative architectures through certain search methods. The best architecture is defined as the one that maximizes generalization accuracy:
$[1] \triangleq \mathop{\arg\max}_{i \in \mathcal{P}} \mathbb{E}(\text{ACC}_i)$,
where $\text{ACC}_i$ denotes the accuracy of alternative $i$. The expectation is theoretically taken over the probability space of all unseen data points, which is impossible to compute exactly, so we estimate it using Monte Carlo simulation over the test dataset.
Therefore, NAS can be viewed as a discrete simulation optimization problem, where the simulation model is a neural network. Testing one architecture on a batch of test data corresponds to one simulation run, and the resulting accuracy is treated as the simulation output.
Given the limited size of the test dataset, which corresponds to a finite amount of simulation resources, we formulate NAS as a fixed-budget R\&S problem.
We compare the following algorithms: OCBA; CBA (correlated budget allocation) from \cite{fu2007simulation}, which extends OCBA by accounting for correlations between alternatives; OCBA and CBA combined with P3C, denoted as P3C-OCBA and P3C-CBA; FBKT from \cite{hong2022solving}; and the naive equal allocation (EA).
Additionally, standard OCBA, CBA, FBKT, and EA are also implemented in a parallel computing environment using the traditional ``divide and conquer'' strategy, as described in Section~\ref{chap_3_P3C}.

\textit{NAS setting.} We conduct experiments on the \textit{CIFAR-10} dataset, which consists of 50,000 training images and 10,000 test images for image classification. The architecture search phase is implemented in PyTorch using the state-of-the-art \textit{Single-Path One-Shot} (SPOS) method \citep{guo2020single}. The search space is a single-path supernet composed of 20 choice blocks connected in series, each with 4 choices. We set the number of alternative architectures to \(p = 10^5\).

\textit{R\&S setting.} The R\&S phase is conducted under the same parallel computing environment as described in Section \ref{chap7.2}. After obtaining $10^5$ alternative architectures, we use {P3C-OCBA, OCBA, P3C-CBA, CBA, FBKT and EA} to allocate computational resources for performance evaluation and select the best architecture. The R\&S performance is evaluated using \(\text{PCS}_{\text{trad}}\). One simulation observation corresponds to testing an architecture on a batch of 32 images. Therefore, the maximum sample size for each alternative is $\frac{10000}{32}\approx312$. The initialization sample size is set to $N_0=20$. The $\text{PCS}_{\text{trad}}$ values are directly estimated based on 1000 independent macro replications. The true best $[1]$ (invisible to the users and used solely for the final $\text{PCS}_{\text{trad}}$ calculation) is estimated by selecting the alternative with the highest accuracy on the full test set. Moreover, in the Stage 2 of ``divide and conquer" and P3C, the $\text{PCS}_{\text{trad}}$ values vary across different processors. The overall $\text{PCS}_{\text{trad}}$ for Stage 2 is approximated by a weighted average based on the number of alternatives on each processor.

\begin{figure}[t]
    \centering
    \includegraphics[width=0.7\textwidth]{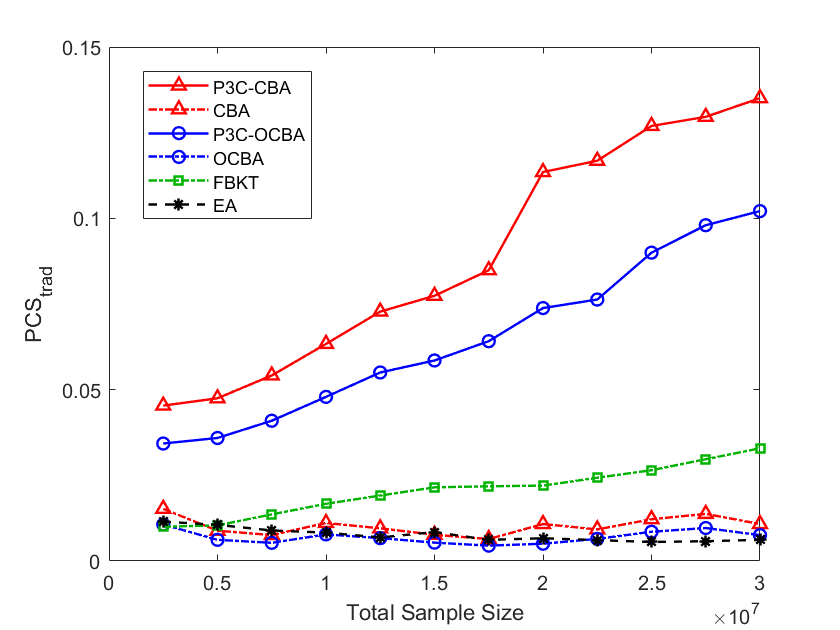}
    \caption{
    Comparison of R\&S procedures under different total sample sizes in the NAS experiment.
    The vertical axis reports the traditional probability of correct selection, $\mathrm{PCS}_{\mathrm{trad}}$,
    while the horizontal axis shows the total sample size (scaled by $10^7$).  
    }
    \label{fig:nas_pcs}
\end{figure}

\begin{table}[t]
    \centering
    \caption{
    Wall-clock running time (in seconds) of different procedures under varying total sample sizes in the NAS experiment.
    }
    \label{tab:nas_time}
    \renewcommand{\arraystretch}{1.05}
    \setlength{\tabcolsep}{5pt}
    \begin{tabular}{c|ccccc}
        \midrule\midrule
        \diagbox[width=20em,height=2.5em]{}{\textbf{Sample Size $(\times10^7)$}} 
        & \textbf{1.0} & \textbf{1.5} & \textbf{2.0} & \textbf{2.5} & \textbf{3.0} \\ 
        \midrule
        \textbf{Clustering Time}  & 10.2 & 10.0 & 10.1 & 10.0 & 10.2 \\ 
        \textbf{CBA}              & 29.9 & 31.2 & 33.7 & 35.6 & 36.7 \\
        \textbf{P3C-CBA}          & 42.1 & 43.4 & 46.4 & 46.2 & 47.1 \\
        \textbf{OCBA}             & 29.0 & 29.4 & 33.0 & 34.7 & 35.8 \\
        \textbf{P3C-OCBA}         & 39.1 & 39.2 & 43.0 & 45.2 & 46.0 \\ 
        \textbf{FBKT}             & 134.1 & 135.2 & 135.4 & 138.2 & 140.9 \\
        \midrule\midrule
    \end{tabular}
\end{table}

{Figure~\ref{fig:nas_pcs} presents the experimental results under different total sample sizes. This is a large-scale R\&S problem with limited simulation resources. As shown in the figure, \(\text{PCS}_{\text{trad}}\) remains extremely low under classic strategies such as OCBA and CBA. }FBKT performs slightly better but still remains below 0.05. However, with P3C, both CBA and OCBA improve by more than an order of magnitude and surpass FBKT, demonstrating the effectiveness of P3C in large-scale fixed-budget R\&S. Notably, P3C-CBA performs slightly better than P3C-OCBA by leveraging correlation information in the sampling policy.

Table~\ref{tab:nas_time} shows the wall-clock time for different algorithms (excluding the lengthy evaluation on the test data and reporting only the time spent on R\&S). Since the total number of alternatives is fixed, the clustering time remains around 10 seconds. The additional time for P3C-CBA and P3C-OCBA, compared to their non-P3C counterparts, is primarily due to the clustering step. FBKT, a dynamic procedure requiring multiple rounds of comparisons, incurs significantly higher computational time.

%
%
%

\section{Concluding Remarks}
In this paper, both theoretical and empirical results indicate that leveraging correlation is an effective way to improve sample efficiency in large-scale R\&S. The P3C procedure presented in this paper is a ready-to-use tool that achieves significant improvements without requiring excessively high clustering accuracy; it only needs to identify which alternatives are highly correlated.

%
%


\bibliographystyle{informs2014} 
\bibliography{ref} 

\end{document}